\documentclass{template/ifacconf}

\usepackage{graphicx, import}      
\usepackage{natbib}        

\usepackage{amsmath, amsfonts, amssymb}
\usepackage[c1,short]{optidef}
\usepackage[utf8]{inputenc}
\DeclareUnicodeCharacter{2212}{$-$}
\usepackage[inline]{enumitem}
\usepackage{xspace}
\usepackage{epstopdf}
\usepackage{lipsum}

\usepackage[dvipsnames]{xcolor}
\usepackage[deletedmarkup=sout,authormarkup=superscript]{changes}
\definechangesauthor[name={Samuel}, color=NavyBlue]{SB}
\definechangesauthor[name={Dominic}, color=RedViolet]{DLM}
\definechangesauthor[name={Alisa}, color=Orange]{AR}
\definechangesauthor[name={ToDo}, color=Green]{TODO}
\definechangesauthor[name={Stefan}, color=Magenta]{STEF}

\def\draft{10}
\usepackage{datetime2}

\begin{document}
\begin{frontmatter}

	\title{Drone-based Volume Estimation in Indoor Environments}

	\thanks[footnoteinfo]{\if \draft
Version compiled: \today~\DTMcurrenttime.
\fi
This research has been supported by NCCR Automation, a National Centre of Competence in Research, funded by the Swiss National Science Foundation (grant number 180545), and by the Swiss Innovation Agency (Innosuisse) via grant number 52481.
}

\author[First,Second]{Samuel Balula} 
\author[First]{Dominic Liao-McPherson}
\author[Third]{Stefan Stevšić}
\author[First,Second]{Alisa Rupenyan}
\author[First]{John Lygeros}

	\address[First]{Automatic Control Laboratory, ETH Zürich, 
	Switzerland \\
	(e-mail: \{sbalula, dliaomc, ralisa, lygeros\}@control.ee.ethz.ch).}
	\address[Second]{inspire AG, 
	Zürich, Switzerland}
	\address[Third]{Tinamu Labs, Zürich, Switzerland \\ (e-mail: stefan@tinamu-labs.com)}
\begin{abstract}                
Volume estimation in large indoor spaces is an important challenge in robotic inspection of industrial warehouses. We propose an approach for volume estimation for autonomous systems using visual features for indoor localization and surface reconstruction from 2D-LiDAR measurements.
A Gaussian Process-based model incorporates information collected from measurements given statistical prior information about the terrain, from which the volume estimate is computed.
Our algorithm finds feasible trajectories which minimize the uncertainty of the volume estimate.
We show results in simulation for the surface reconstruction and volume estimate of topographic data.


\end{abstract}


\end{frontmatter}
\newcommand{\ConfigurationSpace}{\mathcal{C}}
\newcommand{\PlanningSpace}{\mathcal{C_{\mathrm{free}}}}
\newcommand{\Volume}{\mathcal{V}}
\newcommand{\Measurements}{\mathbb{M}}
\newcommand{\Reals}{\mathbb{R}}
\newcommand{\World}{\mathcal{W}}
\newcommand{\Xnav}{X_\mathrm{nav}}
\newcommand{\Xfull}{X_\mathrm{full}}
\newcommand{\tp}{^\mathsf{T}}
\newcommand{\inv}{^{-1}}
\newcommand{\Tr}{Tr}
\newcommand{\cov}{\Sigma}
\newcommand{\mean}{M}
\newcommand{\innovation}{\tilde{y}}
\newcommand{\info}{\Theta}
\newcommand{\state}{\chi}
\newcommand{\fullstate}{\state_\mathrm{full}}
\newcommand{\Input}{u}
\newcommand{\iteration}{j}
\newcommand{\Kalman}{\mathbb{K}}
\newcommand{\tractor}{square wave pattern\xspace}
\newcommand{\zm}{z^m}

\section{Introduction}
\label{sec:introduction}
Autonomous robotic platforms are increasingly used for data collection, for example in structural inspection \cite{almadhoun2016survey}, agriculture surveillance, search and rescue, and industrial environments.
In industrial warehouses it is common to store raw materials as stockpiles, and determining the current amount of material in stock is of paramount importance for logistics. However, to the best of our knowledge, the task of estimating the volume with an automated robotic platform has not been addressed. On the contrary, current business practice is to take differential measurements of the volume added or removed, which is prone to drift over time, and on periodic inspections from experts, which are costly and inaccurate.
A quadcopter equipped with an adequate suite of sensors could be used for this purpose, since it can fly above the pile of bulk material taking advantage of its maneuverability to take measurements from poses otherwise unreachable while avoiding obstacles in cluttered environments.
Estimating the volume using autonomous quadcoptor in indoor environment imposes several requirements, e.g.:
\begin{enumerate*}
	\item indoor localization, where the location uncertainty depends on the drone state,
	\item accurate measurements in environments with uneven light and dust,
	\item an efficient surface reconstruction method, able to cope with large amounts of data, and
	\item path planning, due to the limited time budget to fly the drone and the presence of obstacles.
\end{enumerate*}

\if\draft \subsection{Literature review} \fi
\if\draft \subsubsection{Indoor localization} \fi
Localizing a mobile platform in a GPS-deprived environment with a known map can be achieved using information obtained from dead-reckoning, infrared, radio or sound-based distance measurements, visual information using a motion capture system, or from an onboard camera. In this work localization is inferred from an onboard camera, detecting a set of fixed, previously mapped features, given the lower overall system cost, flexibility and taken into account the accuracy requirements.

\if\draft \subsubsection{LiDAR measurement} \fi
Surface reconstruction can be performed from images, with photogrammetry methods such as structure from motion (SfM)~(\cite{newcombe2011kinectfusion}). This methods are however problematic for objects with homogeneous surfaces or improper lighting. An alternative way is via active laser scanners (LiDAR)
which project a laser beam and measure the time of flight of the reflected light. This sensors are precise and are not affected by the effect of scale uncertainty present in vision based measurements. Furthermore 2D LiDAR systems are lighter than their 3D counterparts, allowing the use of more maneuverable quadrotors.
There are commercial examples of drone-based solutions that use this method in outdoor environments~(\cite{dronedeploy}).
\if\draft \subsubsection{Surface model} \fi
\if\draft \subsubsection{Planning} \fi
The reconstruction quality depends to a large degree on the availability and quality of measurements. Classic approaches for quality-driven and automated 3D scanning use volumetric (\cite{khalfaoui2013efficient}) and Poisson mesh-based metric.
Algorithms also been proposed for multi-view stereo reconstruction (\cite{hepp2018plan3d}), defining heuristics to decide on the utility of the next measurements and optimize set viewpoints based on initial scans. 
Alternatively, 2D laser scanner has been used for 3D mapping in \cite{elasticlidar, loam, loamsimilar}, often mounting the 2D scanner on a rotating motor to emulate 3D LiDAR properties.

In this paper, we consider the problem of estimating the volume of material within a given domain using a drone mounted 2D LiDAR unit operating in an indoor environment, leveraging information about the surface.
This setting has challenges unique to GPS denied environments, notably the uncertainty in the localization depends on the position of the drone, which must follow trajectories that keep enough features in view to maintain localization accuracy.	
In \cite{hollinger2012uncertainty} the authors consider visual inspection of ship hulls using underwater vehicles. In \cite{zhu2021online} the authors consider inspection of 3D object, and \cite{popovic2020informative} deals with a similar problem setting, but localization uncertainty is not taken into account. This work differs since it deals with indoor environments and carefully considers the uncertainty of the measurements.
 Our contributions are threefold: 
\begin{enumerate*}
\item we derive measurement models and uncertainty estimates for the camera-based localization scheme and the LIDAR system used to measure the surface;
\item we propose a scalable methodology for estimating the volume of material based on LIDAR measurements and qualifying the uncertainty of our estimate; 
\item we propose a preliminary informative path planning method that greedily minimizes the uncertainty in the volume estimate.
\end{enumerate*}

\section{Problem Statement}
\label{sec:problem-statement}

We consider the problem of estimating the volume of a pile of bulk material inside a region of interest using a quadrotor-based mobile sensor.
Let $h(x) = h(x_1,x_2)$ be the true surface function of the height of the pile, defined in the domain of interest $\mathcal{D} = [{x_1^-}, {x_1^+}]\times[{x_2^-},{x_2^+}]$.
The volume of the pile is 
\begin{equation}
	\Volume = \iint_\mathcal{D} h(x) dx,
\end{equation}
and the dynamics of the quadcopter are given by
\begin{subequations}
	\begin{align}
		\dot{\state}_\mathrm{full} (t) = &~ f(\fullstate(t),u(t)),\\
		u(t) = &~g(r(t), \hat \state_\mathrm{full}(t))
	\end{align}
\end{subequations}
where $\fullstate(t) = (p,\theta, v, \omega, b):[0,T] \rightarrow \Reals^{13}$ is the drone state, consisting of its position $p$, orientation $\theta$, velocity $v$ and angular velocity $\omega$, as well as the battery state of charge $b$,
$u(t):[0,T] \rightarrow \Reals^{4}$ are the rotor voltages,
$g$ is a feedback controller that stabilizes the system to a commanded position and yaw setpoint $r$, and
$\hat \state_\mathrm{full}(t)$ is an estimate of the current state of the drone.
In this work we assume that the low-level controller can follow the reference closely, such that ${\fullstate}_i(t) \approx r_i(t), i\in\{1,\dots,4\}$ if the time derivatives of $r(t)$ up to order four (velocity, acceleration, jerk and snap) are within specified bounds, derived from actuator limits $\mathcal{U}$ \cite{drone-model}.
\begin{figure}
\begin{center}
	\includegraphics[width=\columnwidth]{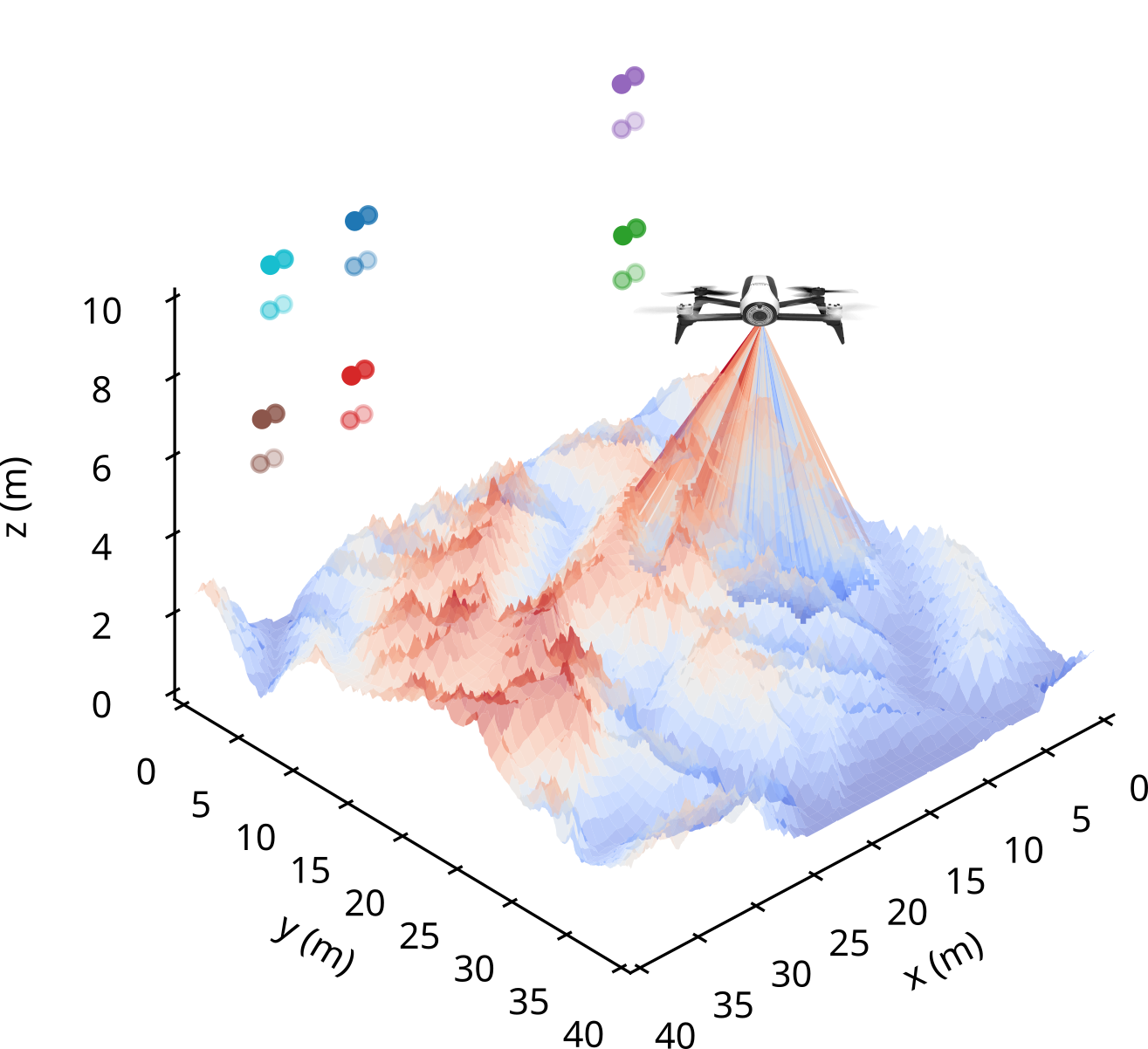}
	\caption{A sample of the scaled topographic data used in simulation. The color circles on the back plane are previously mapped visual features used for localization. The drone represented is not to scale.
	}
\label{fig:alps}
\end{center}
\end{figure}

\if\draft \section{LiDAR model} \fi
The quadcopter is equipped with a 2D LiDAR, measuring the radial distance $d_l$ from the sensor to the surface of the pile, that is used to estimate the volume of the pile based on a reconstruction of the surface. 
%
%
%
The drone is also equipped with a camera and computer vision algorithms that allow it to detect and identify features placed in the environment at known locations. These features are used to localize the drone which must keep a minimum number of features within its field of view at all times.

\if\draft \section{Define objective} \fi
We have now all the ingredients needed to formulate our volume estimation problem. We aim to find a reference trajectory $r(t):[0,T]\rightarrow\PlanningSpace \subseteq \Reals^3\times[0,2\pi[$,
where $T$ is the trajectory time and
$\PlanningSpace$ is the set of 3D positions and yaw where the drone is allowed to fly and is able to detect enough features to localize itself,
that minimizes the volume estimate uncertainty,

\begin{mini!}[2]
	{r(t)}{ \mathbb{V}(\Volume(T))}{\label{main-optimization-problem}}{}
	\addConstraint{ \dot{\state}_\mathrm{full}(t) = }{f(\fullstate(t),u(t))}{}
	\addConstraint{ u(t) = }{g(r(t),\hat \state_\mathrm{full}(t))}{}
	\addConstraint{ \fullstate(t) \in}{\PlanningSpace}{}
	\addConstraint{ u(t) \in}{\mathcal{U}}{},
	\end{mini!}
where $\mathbb{V}$ is the variance operator. This is a hard problem as the uncertainty in the volume estimate is a consequence of uncertainty in the LiDAR measurements which in turn depends on the uncertainty in the camera-based localization system. In the subsequent sections we derive detailed measurement models for the localization system (Section \ref{sec:localization}) and the LIDAR, propose a method for fusing them to estimate the volume (Section \ref{sec:approach}) and propose a greedy algorithm to approximate a solution of \eqref{main-optimization-problem} in Section \ref{sec:planning}.

\section{Localization}
\label{sec:localization}
We are primarily interested in operations in indoor GPS denied environments. Our drone is equipped with an IMU and a front facing camera, which is used to localize the drone based on markers placed in known locations in the environment.
Both markers and natural features are detected using computer vision algorithms (\cite{garrido2014automatic, Koray2009Monocular}) which also have been demonstrated by Tinamu Labs (\cite{jiang2022online}).
%
We require that enough previously mapped features are visible at any given point of the flight to assure that the state can always be determined, independently from other sensory information.
\begin{figure}
\begin{center}
	\includegraphics[width=.6\columnwidth]{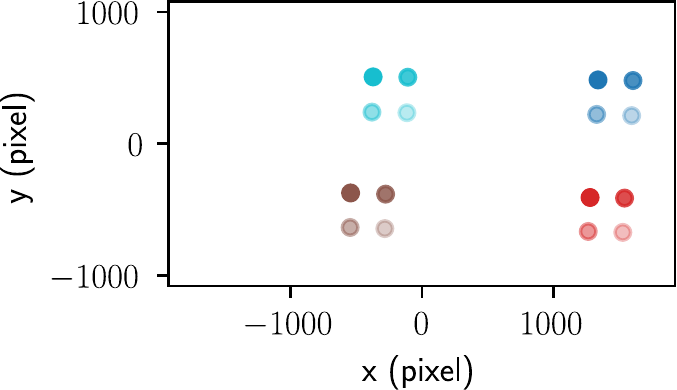}
	\caption{An example of the image in the camera plane, showing the detected points $\overline{P_{im}}$ which are used to infer localization.	
	}
\label{fig:camera}
\end{center}
\end{figure}
Let $\overline{P_{im}} = \{{P_{im}}_k\}_{k=1}^N \subset \Reals^2$ be the set of $N$ features detected and identified in the image plane of the onboard camera, as exemplified in Figure \ref{fig:camera}.
We can compute the coordinates in the image frame $P_{im}$ from the global frame $P_g$ as a function of the position and orientation of the drone $\state=\{{\state_\mathrm{full}}_i\}_{i=1}^6$
\begin{equation}
	P_{im}(\state) = C_{im}^c T_c^g(\state) P_g,
\end{equation}
where $T_c^g(\state)$ is the geometric transform from the global to the camera frame, and $C_{im}^c$ the transform from camera frame to image plane given by the camera model.
We can solve a nonlinear least squares problem to find an estimate of $\state$
\begin{equation}\label{nlleastsquares}
	\hat \state = \underset{\state}{\arg \min} \sum_{k=1}^N \|P_{im}(\state) - \overline{P_{im}}\|_2^2.
\end{equation}

Linearizing the nonlinear least squares problem around the solution, 
for normally distributed measurement errors, the estimate $\hat \state$ is also normally distributed with variance 
$	\mathbb{V}(\hat \state) = \sigma^2 (J\tp J)\inv, $
where $\sigma$ is the standard deviation of a measurement in the image plane, and $J = \nabla P_{im}(\state)|_{\state=\hat \state}$ is the Jacobian.
%
Assuming $\hat \state \approx \state$, we can compute the variance of the state estimate $\Sigma^\state$ as a function of the drone coordinates
\begin{equation}
	\Sigma^\state(\state) = \mathbb{V}(\hat \state)|_{\hat \state = \state}.
\end{equation}

Let $q_{pos}(\Sigma^\state(\state)) = 1/\sqrt{\Tr(\Sigma^\state)}$ 
be a function mapping the covariance matrix to a scalar \emph{quality of position fix} metric. This metric is used to define the 
%
\if\draft \subsection{State space constraints} \fi
feasible flying domain $\PlanningSpace$ as
\begin{equation}
	\PlanningSpace = \{\state \in \ConfigurationSpace : q_\mathrm{pos}(\Sigma^\state(\state)) > \tau ~|~ \World \}
\end{equation}
where $\state$ is the drone state,
$\ConfigurationSpace$ is the configuration space,
$\tau$ is the minimum quality of fix threshold,
and $\World$ the available information about the visual features positions.

The IMU and position measurements derived from the vision system are fused using an extended Kalman filter \cite{ekf} to produce an estimate of the full state of the drone.
The full estimate is used by the inner loop controller $g$ while the planning algorithm only use the position and yaw localization.

\section{Volume estimation}
\label{sec:approach}
\if\draft \subsection{State estimation} \fi 
%
%

\begin{figure}
\begin{center}
	\def\svgwidth{.5\columnwidth}
	\immediate\write18{inkscape -z -D --file=img/DroneLidarHit.svg --export-pdf=img/DroneLidarHit.pdf --export-latex}
	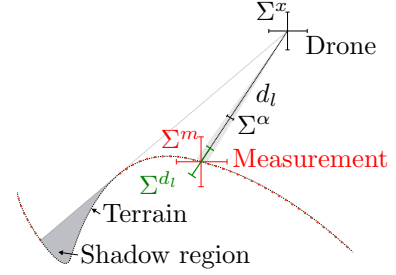%
	\caption{The 2D LiDAR measures the distance from the drone to the terrain $d_l$. The uncertainty in the drone position and orientation $\Sigma^\state$, LiDAR angle $\Sigma^\alpha$ and measurement distance $\Sigma^{d_l}$ are propagated to obtain the covariance of the measurement $\Sigma^m$.
	Due to features of the terrain some areas might not be visible from current pose, marker as the shadow region in the bottom right.
\label{fig:DroneLidarHit}
	}\end{center}
\end{figure}

Now that we know the position and orientation of the drone, as well as their uncertainties we consider the LiDAR model.
The sensor rotates in a plane with constant angular velocity, collecting approximately $10\mathrm{K}$ distance samples per second, at discrete angles as illustrated in Figures~\ref{fig:alps} and~\ref{fig:DroneLidarHit}.
We can compute the hit point coordinates of the LiDAR measurements with the geometric transformation
\begin{equation}
	z^m(\state, \alpha) = T_g^l(\state, \alpha) d_l,
\end{equation}
where
$T_g^l(\state,\alpha)$ is the geometric transformation converting points in the LiDAR scan line frame to global coordinates,
$\alpha$ is the angle of the sensor,
$d_l$ is the measured distance, and
$\zm\in\Reals^3$ are the coordinates of the hit point.
Propagating the uncertainty of each of the variables we obtain an estimate of the covariance of the measurement $\zm$, where again we use the Jacobian of the geometric transformation
\begin{equation}
	\begin{aligned}
		\Sigma^m &=  \nabla_\state T_g^l(\state, \alpha) d_l ~ \Sigma^\state ~ (\nabla_\state T_g^l(\state, \alpha) d_l)\tp\\
		& + \nabla_\alpha T_g^l(\state, \alpha) d_l ~ \Sigma^\alpha ~ (\nabla_\alpha T_g^l(\state, \alpha) d_l)\tp \\
		& + T_g^l(\state, \alpha)~ \Sigma^{d_l} ~ T_g^l(\state, \alpha)\tp,
	\end{aligned}
	\label{sigma-m}
\end{equation}
where $\Sigma^m$, $\Sigma^\state$, $\Sigma^\alpha$ and $\Sigma^{d_l}$ are the covariance matrices of a measurement, the drone position and orientation, the LiDAR angle, and measured distance, respectively.
Note that due the intrinsic properties of the sensor, only obstacles in a certain distance range can be detected $d_l \in [d_{min}, d_{max}]$.
Figure \ref{fig:DroneLidarHit} shows how the LiDAR scan obtains data about the surface.
\if\draft \subsection{Reducing errors-in-variables to z} \fi
\begin{figure}
\begin{center}
	\def\svgwidth{.9\columnwidth}
	\immediate\write18{inkscape -z -D --file=img/Terrain.svg --export-pdf=img/Terrain.pdf --export-latex}
	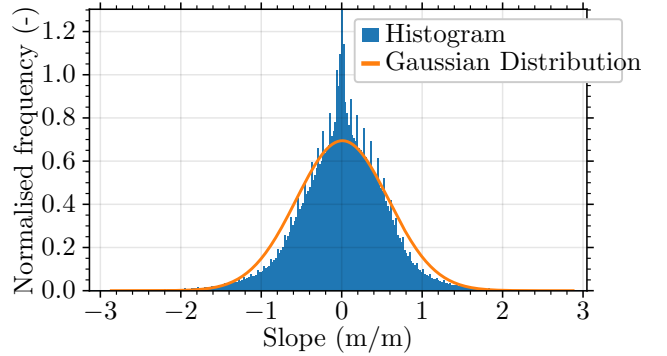%

	\caption{The volume slope distribution can be approximated by a normal distribution $\mathcal{N}(0, \sigma_s)$, and it is a property of the material.} 
	\label{fig:slope}
\end{center}
\end{figure}
We condense the position uncertainty $\Sigma^m$, propagating the errors-in-variables to errors in height, given known statistical information about the slope of the terrain
\begin{equation}
	\Sigma^m_{(z)} = \begin{bmatrix}\sigma_t &\sigma_t &1 \end{bmatrix} \Sigma^m \begin{bmatrix}\sigma_t &\sigma_t &1 \end{bmatrix}\tp,
\end{equation}
where $\sigma_t$ is the standard deviation of a Gaussian distribution fitted to an histogram of the slopes of this terrain, as shown in Figure \ref{fig:slope}.


\subsection{Surface model}
A common way to store the information about a 3D object is to use voxels, a generalization of the concept of pixel in 3D. Variations include sparse octree representations \cite{laine2010efficient}. However, if we assume that the surface can be described with a function from $x, y$ to $h$ we can simplify the representation.
One option is to use kriging-based methods (\cite{krige1951statistical}) such as Gaussian process (GP) (\cite{rasmussen}). This is the method used in ~\cite{popovic2020informative} where an informative path planning framework is proposed.

We parameterize the surface using a grid of heights. Each point of the grid is described by a univariate normal distribution
\begin{equation}
	h_{i,j} \sim \mathcal{N}(\mu_{ij},\sigma_{ij}^2),~i=\{1, \dots, N\}, ~j=\{1, \dots, M\},
	\label{eq:h-diagonal}
\end{equation}
where $N$ and $M$ are the dimensions of the grid. Between the grid points, we represent the surface distribution (the surface height is a random variable due to uncertainty) using a Gaussian Process model where the height grid provides inducing points. 

We use a specific kernel tailored to the physics of our surface. Due to their physical properties, the materials piled up in the region of interest have a volumetric organization which can be described statistically.
The Matérn Kernel is a good choice to estimate the correlation of the heights of two points separated by a distance $s$
\begin{equation}
	k(s) =  \frac{1}{\Gamma(\nu)2^{\nu-1}}\Bigg(
         \frac{\sqrt{2\nu}}{l} s
         \Bigg)^\nu K_\nu\Bigg(
         \frac{\sqrt{2\nu}}{l} s \Bigg),
		 \label{eq:kernel}
\end{equation}
where $l$ is the lengthscale, $\nu$ is a positive parameter controlling the smoothness of the function, and $K_{\nu}$ and $\Gamma_{\nu}$ are Bessel and Gamma functions, respectively.
We describe the surface as a sparse Gaussian Process with fixed inducing points $X$ which are obtained from the height grid,
$f(x) \sim \mathcal{GP} (0, k(X, X'))$.
At an arbitrary set of points $X_* \in \mathcal{D}$, we can predict the expected value and variance of the height of the surface using the equations
\begin{subequations}
	\begin{align}
		\mean^{f_*}_\info = \mathbb{E}[f_*] =&~ K_{X_*X}[K_{XX} + \Sigma^2 I ]^{-1} Z\\
		\begin{split}
			\cov^{f_*}_\info = \mathbb{V}[f_*] =&~ K_{X_*X_*} - \\
					& K_{X_*X}[K_{XX} + \Sigma^2 I]^{-1} K_{XX_*}
		\end{split}
	\end{align}
\end{subequations}
where $\mathbb{E}$ is the expected value,
$\mathbb{V}$ is the variance,
$K_{X_*X}$ is the covariance between the points $X$ and $X_*$ computed with the kernel \eqref{eq:kernel},
where $s$ is the pairwise distance between the points (\cite{rasmussen}),
$Z$ is the vector of the $z$ coordinates of the inducing points, and
$\Sigma$ the vector of their uncertainties.
For a certain kernel lengthscale, the vector $\info \doteq [Z, \Sigma]$ contains all the information needed to make predictions.
For computational reasons, the kernel is made sparse by setting to zero the correlation between points outside of a ball with radius $\gamma l$.
With this model for the surface, we can readily compute the volume, its expected value and variance using the equations
\begin{subequations}
	\begin{align}
		\Volume = &\iint_S f (x_*) dS \\
		\mu^\Volume = & ~ \mathbb{E}(\Volume) \approx  A_\square \sum_{i=1}^N \sum_{j=1}^M \mean^{f_*}_{ij},\label{eq:V-mu} \\
		{(\sigma^\Volume)}^2 = & ~ \mathbb{V}(\Volume) \approx A_\square^2 \sum_{i=1}^N \sum_{j=1}^M \cov^{f_*}_{ij}, \label{eq:V-sigma}
	\end{align}
\end{subequations}
where $A_\square$ is the area of the base of each square cuboid.

\subsection{Update}
When a new measurement is obtained we first propagate it to the grid points. This is accomplished using the kernel
${\Sigma^m_{(z)}} + \sigma_t^2 (e^{s/l}-1)$,
where $\sigma_t$ is the standard deviation of the normal distribution that approximates the slope distribution,
$s$ is the distance between the measurement point and the grid coordinates, and
$l$ is the lenghtscale used in \eqref{eq:kernel}. 
We then use a Kalman filter to update the parameters of the height model \eqref{eq:h-diagonal}
\begin{equation}
	\begin{aligned}
		\mu_k		&= (I - K_k) \mu_{k-1} + K_k z_k \\
		\sigma_k	&= (I - K_k) \sigma_{k-1} \\
		K_k			&= \frac{\sigma_{k-1}^2}{\sigma_{k-1}^2 + ({\Sigma^m_{(z)}}_k + \sigma_t^2 (e^{s/l}-1))}
	\end{aligned}
\end{equation}
where
$\mu_{k-1}$, $\mu_k$, $\sigma_{k-1}$, and $\sigma_{k}$ are the mean and standard deviation of each point of the grid, before and after the update step, 
$z_k$ and ${\Sigma^m_{(z)}}_k$ are the measurement value and covariance from \eqref{sigma-m}.
$K_k$ is the Kalman gain.

\section{Planning}
\label{sec:planning}
\if\draft \subsection{Data structure used for planning} \fi
To estimate the volume of the surface, the robot needs to collect information of the whole surface, moving on a path suitable for this purpose. Two main approaches exist in path planning, that achieve this goal. Coverage path planning \cite{galceran2013survey} is a method to design paths that visit all points of interest while avoiding obstacles. Typical cost functions minimize the length of the path, whereas with informative path planning \cite{stache2021adaptive} the objective is to maximize the amount of information in a feasible path. An overview of path planning algorithms can be found in \cite{lavalle2006planning}.

For the purposes of planning we
\begin{enumerate*}
	\item assume a constant nominal surface height $h_0$, which allows us to compute $d_l$;
	\item decouple the motion of the drone and the rotating LiDAR scan, since the latter is at least one order of magnitude faster;
	\item neglect the possible effect of shadows, as defined in Figure \ref{fig:DroneLidarHit};
	\item we find a discrete set of waypoints instead of a continuous time trajectory.
\end{enumerate*}
We use a greedy algorithm for planning, where the next point to be picked is the one with that minimizes the volume estimate uncertainty
\begin{argmaxi!}[2]
	{r_{k+1}}{ \| \sigma^\Volume_{k+1}(r_{k+1})\|_2}{}{r^*_{k+1} = }
	\addConstraint{\label{eq:motion} \| r_{k+1} - r_k \|_2 \leq ~}{R}{}
	\addConstraint{ r \in}{\{\PlanningSpace_i\}_{i=1}^4}{},
\end{argmaxi!}
where $\sigma^\Volume$ is computed with \eqref{eq:V-sigma},
$r$ is the reference,
$R$ is the radius of a ball where the next step can lie.
The simulations of the next section consider only the $x$ and $y$ components of $r$, and fix $z$ and the yaw.
The greedy algorithm is suboptimal but fast to evaluate, and allows us to test the surface reconstruction method. Future work will focus on implementing more advanced planning algorithms.

\section{Simulations}
\label{sec:results}
We demonstrate our approach for volume estimation on a topographic map of the Alps mountain range, rich on features, which we scale down by a factor of 1000x.
\begin{figure}
\begin{center}
	\includegraphics[width=.49\columnwidth]{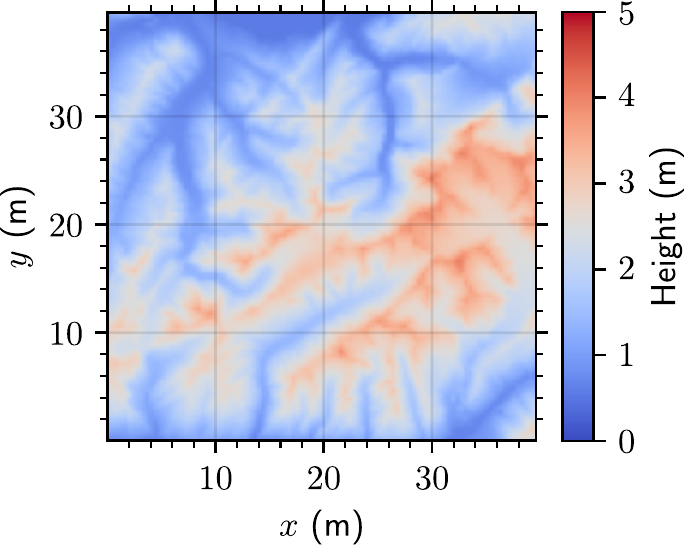} 
		\includegraphics[width=.48\columnwidth]{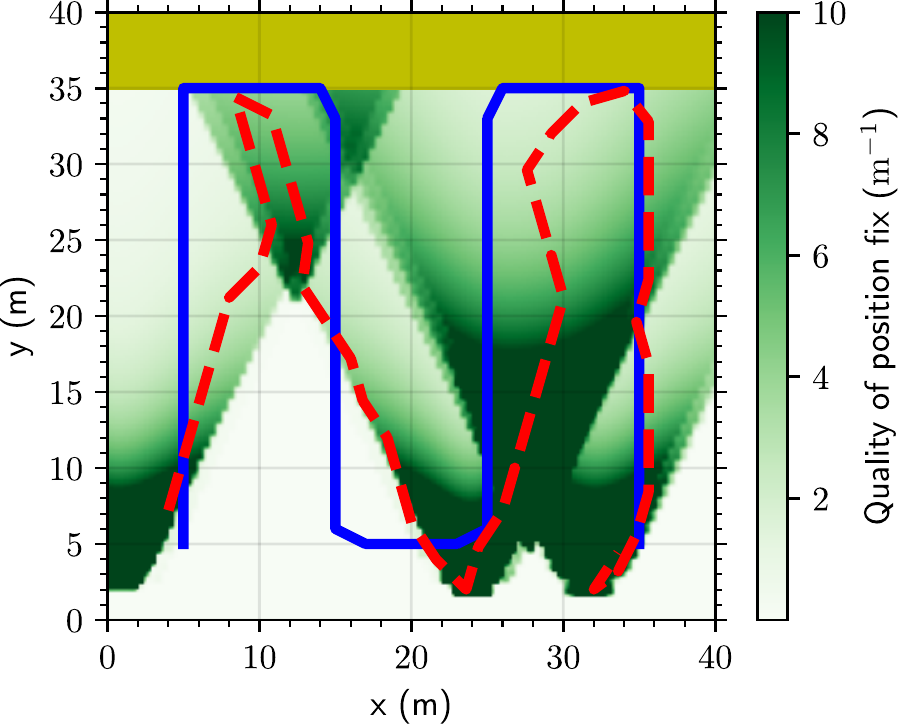} 
		\caption{\label{fig:ground-truth} \label{quality-of-fix}
		The left plot shows the ground truth surface height used in simulation.
		The right plot shows the quality of fix $q_{pos}$ as a function of the $xy$ coordinates, for $z=7\,\mathrm{m}$ and zero yaw.
		Overlayed are the trajectories of the \tractor in solid blue and the greedy algorithm in dashed red.
		The yellow region on the top of the plot indicates constraints in the position coordinates.}  
	\end{center}
\end{figure}
\begin{figure}
\begin{center}
	\includegraphics[trim=35pt 3pt 17pt 3pt, clip, width=.49\columnwidth]{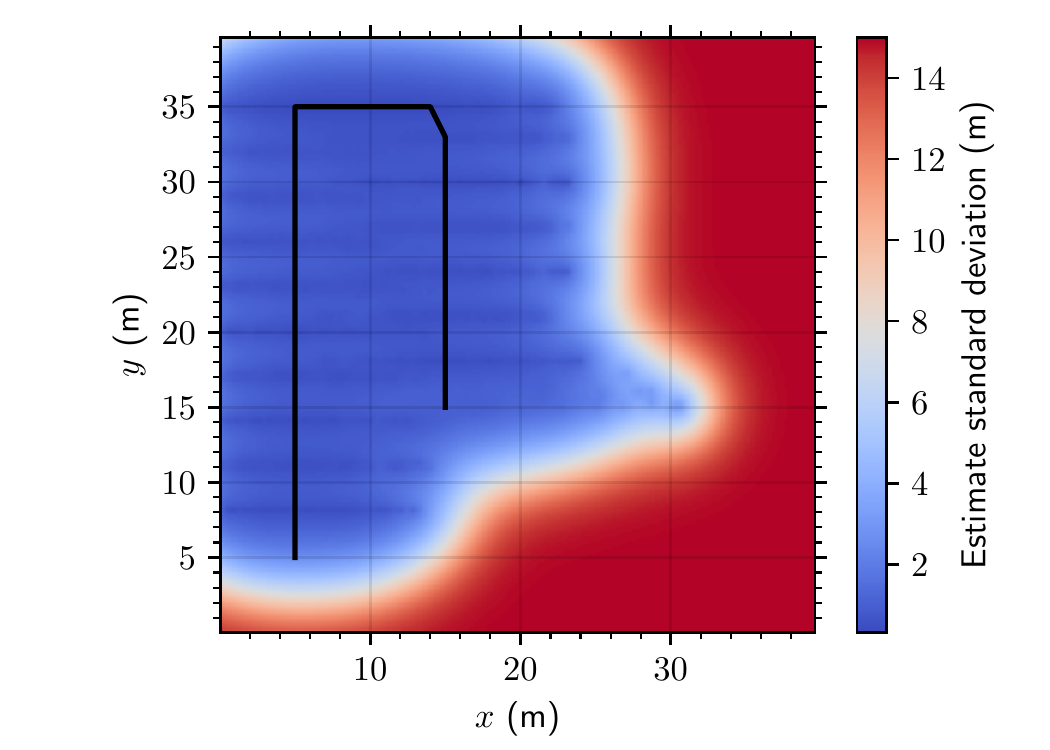} 
	\includegraphics[trim=35pt 3pt 17pt 3pt, clip, width=.49\columnwidth]{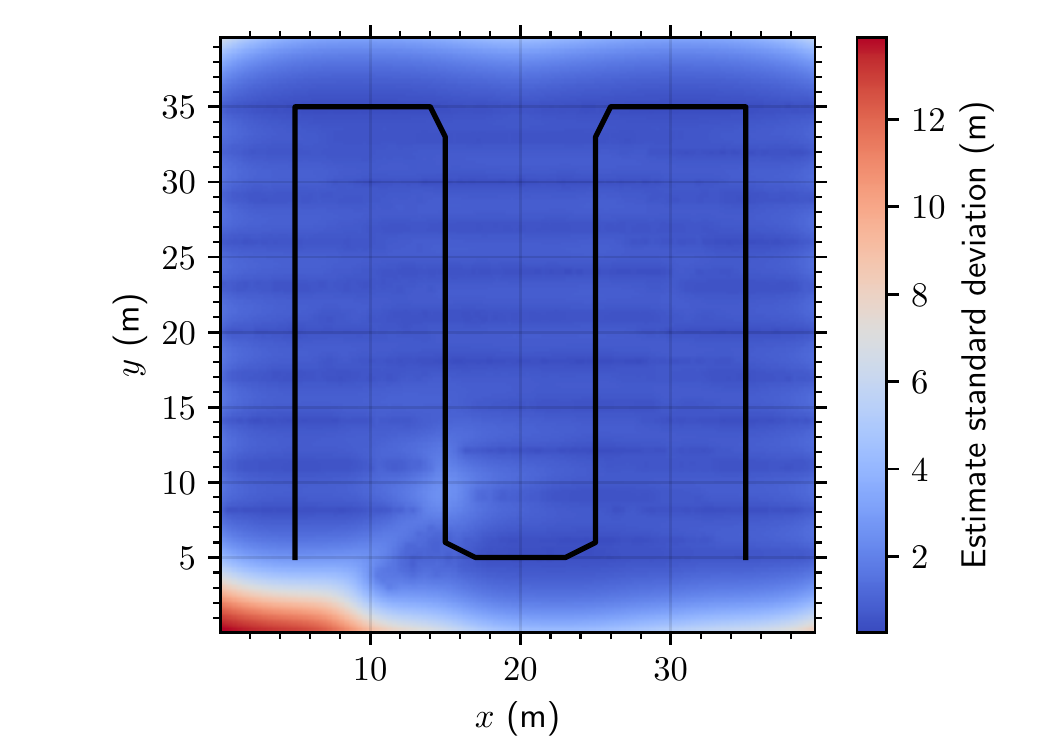} 
	\caption{Uncertainty map of the surface reconstruction with the \tractor after $20$ steps (left plot) and $50$ steps (right plot). \label{fig:tractor-reconstruction-sigma}
	}\end{center}
\begin{center}
	\includegraphics[width=.49\columnwidth]{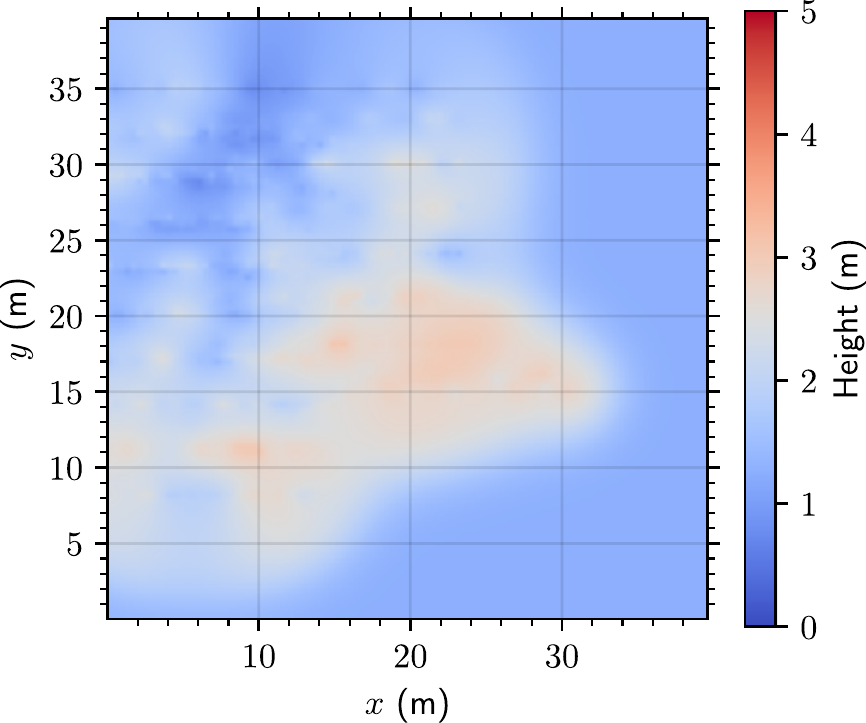} 
	\includegraphics[ width=.49\columnwidth]{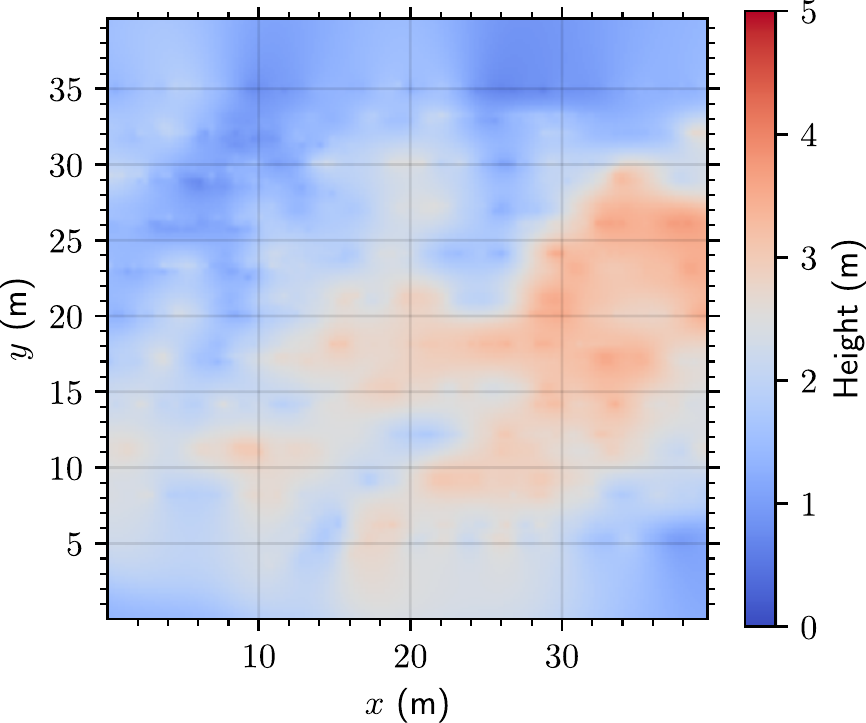} 
	\caption{Surface reconstruction with the \tractor after $20$ steps (left plot) and $50$ steps (right plot).\label{fig:tractor-reconstruction}
	}\end{center}
\end{figure}
\begin{figure}
\begin{center}
	\includegraphics[trim=35pt 3pt 17pt 3pt, clip, width=.49\columnwidth]{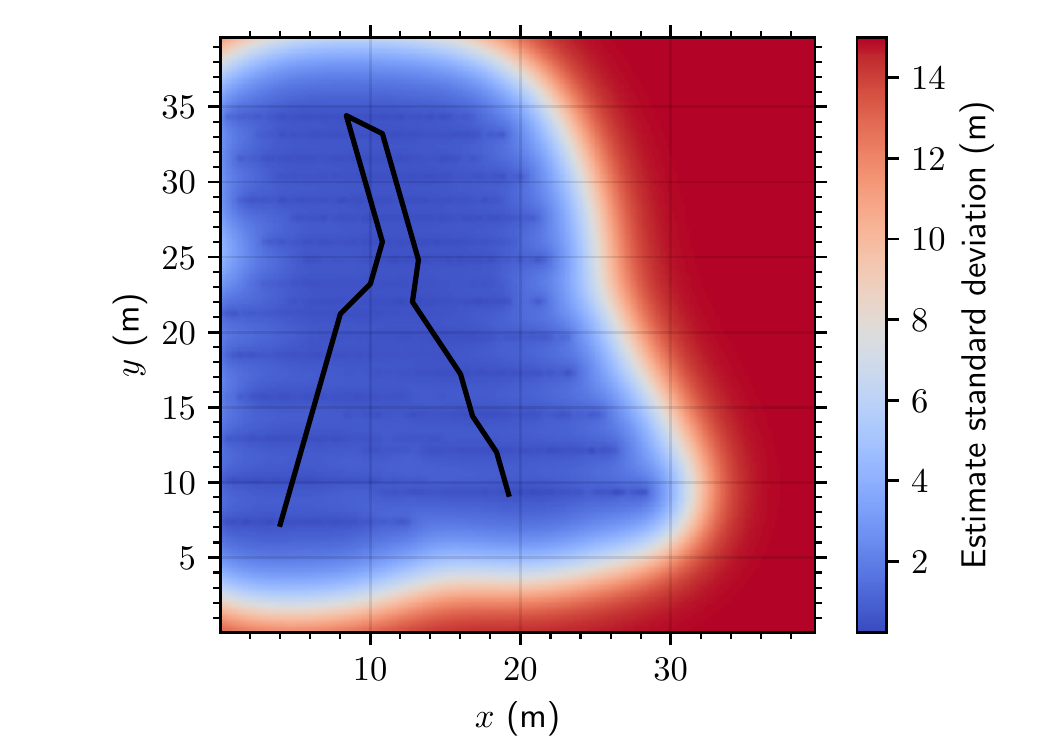} 
	\includegraphics[trim=35pt 3pt 17pt 3pt, clip, width=.49\columnwidth]{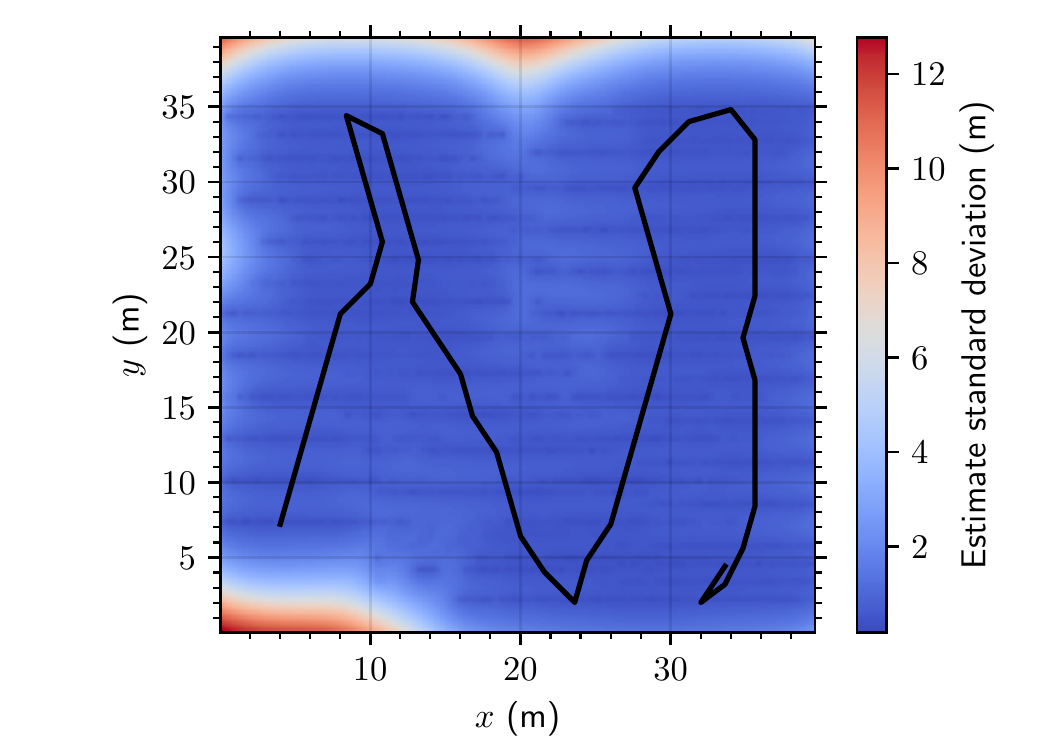} 
	\caption{Uncertainty map of the surface reconstruction with the greedy planner after $20$ steps (left plot) and $50$ steps. \label{fig:greedy-reconstruction-sigma}
	}\end{center}
\begin{center}
	\includegraphics[ width=.49\columnwidth]{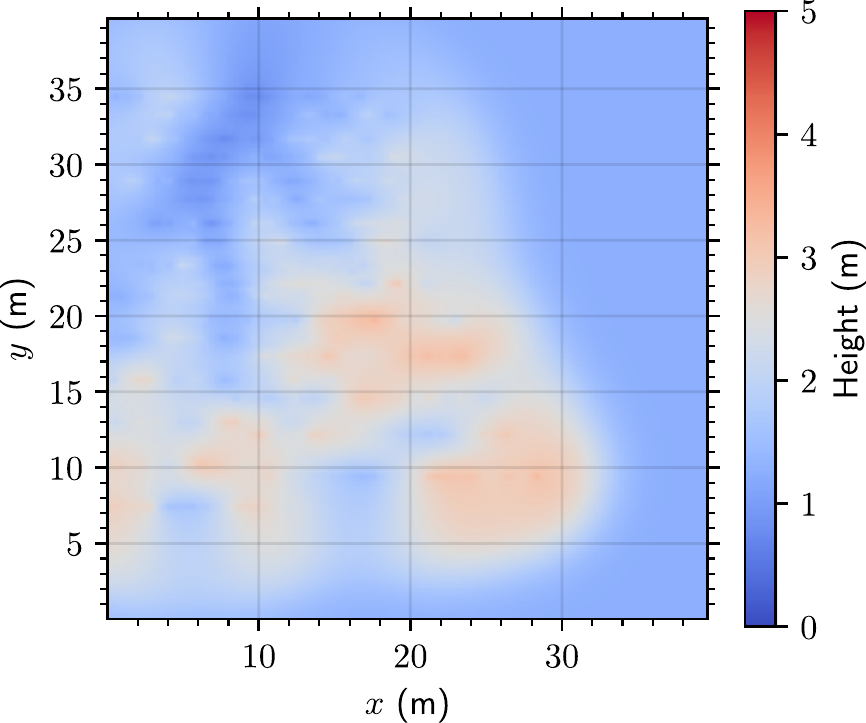} 
	\includegraphics[width=.49\columnwidth]{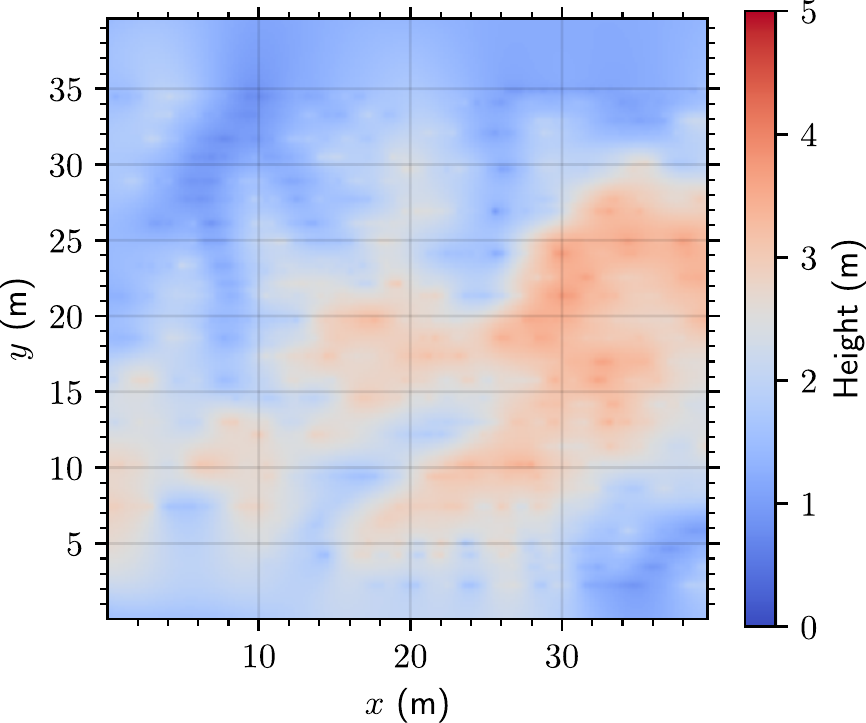} 
	\caption{Surface reconstruction with the greedy planner after $20$ steps (left plot) and $50$ steps (right plot). \label{fig:greedy-reconstruction}
	}\end{center}
\end{figure}
We show simulation results for two trajectories, where the first one is a fixed, manually created \tractor, and the second is the trajectory resulting from applying the path planning algorithm from Section \ref{sec:planning}, in the variables $r_x$ and $r_y$, while keeping $r_z$ and $r_{\mathrm{yaw}}$ fixed.
The disposition of visually identifiable features in the example used in this simulations, shown in Figure \ref{fig:alps}, creates an uneven uncertainty of position map, as shown in the right plot of Figure \ref{quality-of-fix}, where darker colors represent lower uncertainty. 

\begin{figure}
\begin{center}
	\includegraphics[width=.49\columnwidth]{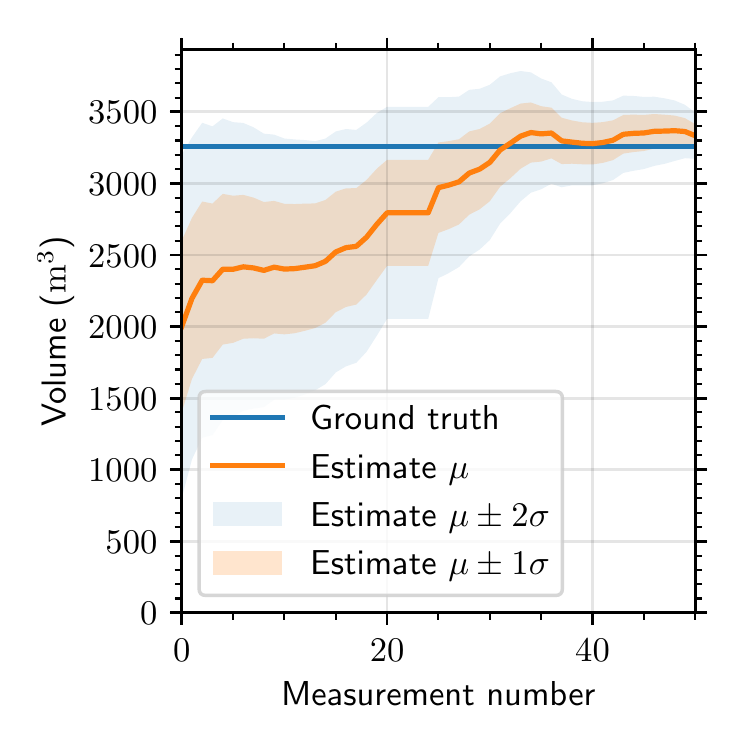} 
	\includegraphics[width=.49\columnwidth]{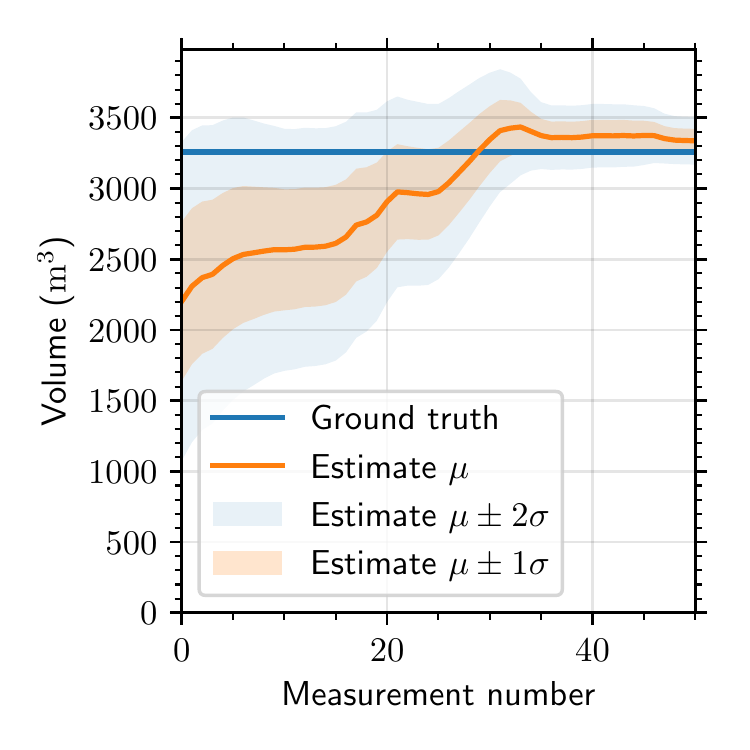} 
	\caption{\label{fig:volume-vs-iteration} Evolution of the volume estimate and its uncertainty along the trajectory. The left plot shows the result for the \tractor and the right plot for the greedy algorithm.
	}\end{center}
\end{figure}

Figure \ref{fig:tractor-reconstruction-sigma} shows the path of the drone and the altitude uncertainty map, and
Figure \ref{fig:tractor-reconstruction} the surface reconstruction map for the fixed \tractor. Traversing the whole region results in reduced uncertainty of surface reconstruction. It is in general not trivial to manually design paths that avoid constraints or regions with insuficient localization quality.
As for the greedy algorithm and its resulting trajectory, we show in
Figure \ref{fig:greedy-reconstruction-sigma} the path of the drone and the altitude uncertainty map, and in
Figure \ref{fig:greedy-reconstruction} the surface reconstruction map.
The simulation results show that a feasible reference trajectory is found that visits most of the region of interest through regions with a high quality of position fix and drives the uncertainty of the volume down to $\sigma^\Volume/\mu^\Volume=2.26 \%$ and a relative error of $2.53\%$ for the greedy algorithm, comparable with $\sigma^\Volume/\mu^\Volume=2.42 \%$ with a relative error of $2.30\%$ for the \tractor.
We speculate that the two methods perform similarly in this example due to the greedy nature of the planning algorithm used, which optimizes only for the next step ahead, thus finding only an approximate solution for the optimization problem defined in \eqref{main-optimization-problem}. Also note that in this preliminary result the yaw and altitude are kept fixed, so the planner has less degrees of freedom to explore. These factors taken together make it difficult to improve on the benchmark \tractor.
Figure \ref{fig:volume-vs-iteration} shows the evolution of the volume estimate as well as the uncertainty as a function of the number of samples collected. The two paths have the same length and number of samples, and the evolution of the volume and its uncertainty is similar for both paths.
Furthermore, the final reconstructions from Figures \ref{fig:tractor-reconstruction} and \ref{fig:greedy-reconstruction} approximate the ground truth shown in Figure \ref{fig:ground-truth} adequately.
In summary the simple greedy planner finds a path that is by all metrics similar to the manually designed \tractor, allowing the automation of the task of designing paths for experimental volume estimation campaigns.

\section{Conclusion}
\label{sec:conclusion}
We develop and implement a framework for volume and uncertainty estimation for an autonomous robotic platform collecting experimental data with a LiDAR scanner under variable position uncertainty.
The surface reconstruction and volume estimation are validated in simulation with a feature rich surface, for two trajectories generated manually and with a simple greedy algorithm.
Future work will focus on the implementation of an improved planning algorithm employing a longer prediction horizon, validation in more complex maps with intricate constraints as well as experimental validation of the method.


\bibstyle{ifacconf.bst}
\bibliography{src/D-bibliography}             

\end{document}